\documentclass[sn-mathphys-num]{sn-jnl}

\usepackage{graphicx}%
\usepackage{multirow}%
\usepackage{amsmath,amssymb,amsfonts}%
\usepackage{amsthm}%
\usepackage{mathrsfs}%
\usepackage[title]{appendix}%
\usepackage{xcolor}%
\usepackage{textcomp}%
\usepackage{manyfoot}%
\usepackage{booktabs}%
\usepackage{algorithm}%
\usepackage{algorithmicx}%
\usepackage{algpseudocode}%
\usepackage{listings}%
\usepackage[T1]{fontenc} 
\usepackage{multibib}
\newcites{appx}{References}

\counterwithin*{figure}{section}
\counterwithin*{table}{section}

\newcommand{\NII}{\mbox{[N\,\textsc{ii}]}}

\theoremstyle{thmstyleone}%
\theoremstyle{thmstyletwo}%

\theoremstyle{thmstylethree}%

\begin{document}
\newcommand{\mnras}{Mon. Not. R. Astron. Soc.}
\newcommand{\araa}{Annu. Rev. Astron. Astrophys.}
\newcommand{\apj}{Astrophys. J.}
\newcommand{\apjl}{Astrophys. J.}
\newcommand{\apjs}{Astrophys. J.}
\newcommand{\aap}{Astron. Astrophys.}
\newcommand{\nat}{Nature}
\newcommand{\pasj}{Publ. Astron. Soc. Jpn}
\newcommand{\pasp}{Publ. Astron. Soc. Pac.}

\title[Bulk Motion of Gas in the Centaurus Cluster]{The Bulk Motion of Gas in the Core of the Centaurus Galaxy Cluster
}


\author[1]{\fnm{XRISM} \sur{collaboration}}



\affil[1]{A list of authors and their affiliations appears at the end of the paper.}




\abstract{
Galaxy clusters contain vast amounts of hot ionised gas known as the intracluster medium (ICM). In relaxed cluster cores, the radiative cooling time of the ICM is shorter than the age of the cluster, but the absence of line emission associated with cooling suggests heating mechanisms that offset the cooling, with feedback from active galactic nuclei (AGNs) being the most likely source \cite{2002MNRAS.332..729C,2007ARA&A..45..117M}. Turbulence and bulk motions, such as the oscillating ("sloshing") motion of the core gas in the cluster potential well, have also been proposed as mechanisms for heat distribution from the outside of the core  \cite{2004ApJ...612L...9F,2010ApJ...717..908Z}. Here we present X-ray spectroscopic observations of the Centaurus galaxy cluster with the XRISM satellite. We find that the hot gas flows along the line of sight relative to the central galaxy, with velocities from $130\rm\: km\: s^{-1}$ to $310\rm\: km\: s^{-1}$ within $\sim 30$~kpc of the centre. This indicates bulk flow consistent with core gas sloshing. While the bulk flow may prevent excessive accumulation of cooled gas at the centre, it could distribute the heat injected by the AGN and bring in thermal energy from the surrounding ICM. The velocity dispersion of the gas is found to be only $\lesssim 120\rm\: km\: s^{-1}$ in the core, even within $\sim 10$~kpc of the AGN. This suggests that the AGN's influence on the surrounding ICM motion is limited in the cluster.
}

\maketitle


As one of the closest cool-core clusters, the Centaurus cluster is an ideal target for studying the complex physics of the ICM. Its central galaxy, NGC~4696 at a distance of 36.8~Mpc \cite{2001MNRAS.327.1004B}, hosts an AGN that has inflated numerous cavities in the surrounding X-ray gas \cite{2016MNRAS.457...82S} (Extended Data Fig.~\ref{fig:cavity}). 
Previous X-ray observations have revealed a wealth of structure in the ICM, including filaments and cold fronts \cite{2016MNRAS.457...82S}. These features are thought to be related to AGN feedback and the sloshing of the core gas. However, the limited spectral resolution of the previous generation of X-ray observatories could not provide hard evidence to support these ideas, especially in terms of dynamics.
We present new observations of the Centaurus cluster core with the X-Ray Imaging and Spectroscopy Mission (XRISM) launched in 2023 \cite{Tashiro2024}. The high spectral resolution, non-dispersive Resolve microcalorimeter on XRISM provides an unprecedented view of the thermodynamics and kinematics of the ICM, allowing the first accurate measurements of the large- and small-scale gas velocities in Centaurus. 
The angular resolution of Resolve ($\sim 14$~kpc for the half power diameter (HPD), at the distance of Centaurus) allows useful spatially resolved spectroscopy.
To unravel the environment around the central AGN, we focus on the central $r\lesssim 30$~kpc region and study the temperature structure, the bulk motions of the gas, and the turbulent velocity. 
Since the power of the AGN, estimated from radio luminosity and cavity size, is smaller than typical for central AGNs in other clusters \cite{2023PASJ...75..925F}, we expect the core of the Centaurus cluster to be less affected by AGN activity.

\begin{figure}[ht]
\centering
\includegraphics[width=\linewidth]{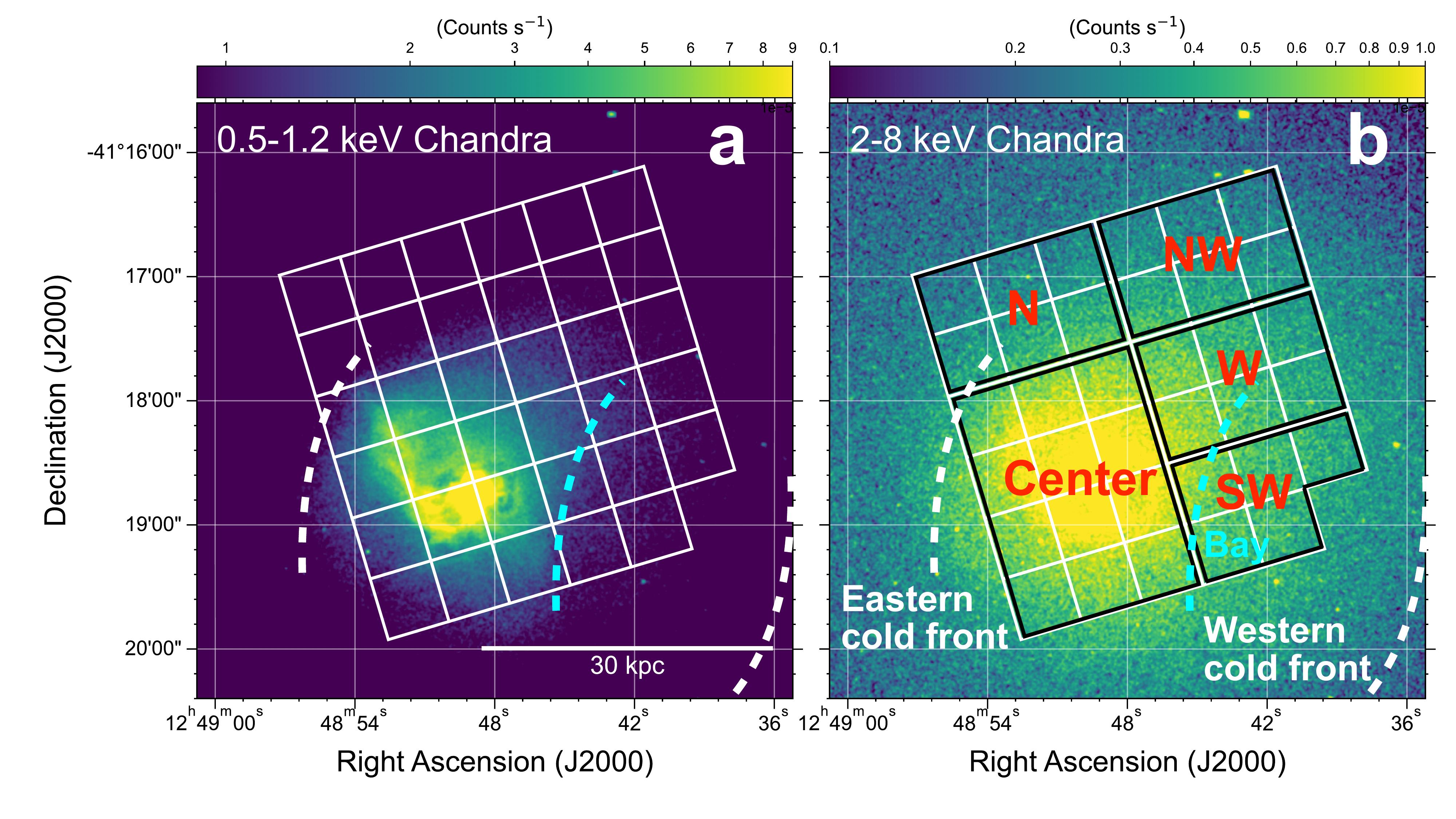}
\caption{{\bf | Resolve field of view (FOV) superimposed on the Chandra X-ray images.} {\bf a.} 0.5-1.2 keV and {\bf b.} 2--8 keV. The red letters correspond to the regions extracted in the spectral analysis. The cyan dashed line indicates the "bay" structure, while the white dashed lines show the eastern and western cold fronts, respectively, as indicated in \cite{2016MNRAS.457...82S}.}\label{fig:image}
\end{figure}

\begin{figure}[ht]
\centering
\includegraphics[width=\linewidth]{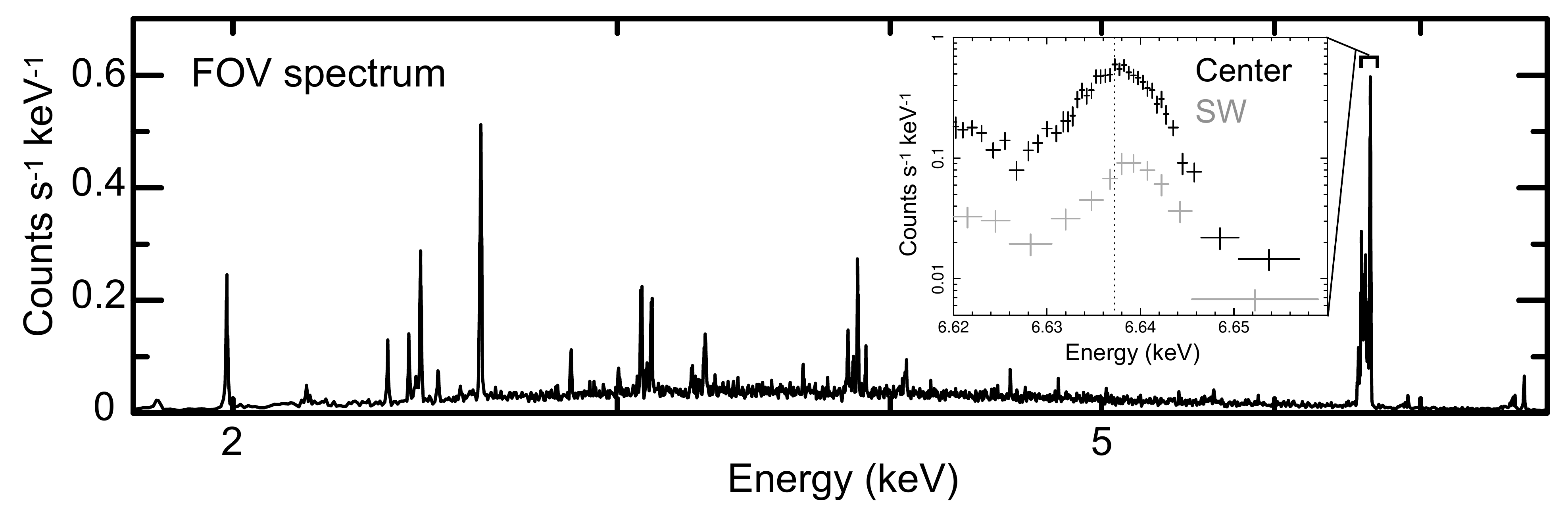}
\caption{{\bf | Resolve spectrum observed for the FOV in the 1.8--8.0 keV band.} In the inset, Resolve spectra around the He-$\alpha$ Fe lines in 6.62--6.66 keV of the Central and SW regions are shown in black and light grey, respectively. The vertical dotted black line indicates the central energy of the redshifted He-$\alpha$ Fe resonance line for the Central region obtained in the SSM analysis. The peak of the SW spectrum is blueshifted with respect to the line.}\label{fig:spec}
\end{figure}

XRISM observed the Centaurus cluster during its performance verification phase from December 2023 to January 2024, with a total exposure time of 284.7 ksec.
The aim point for the Centaurus cluster was located at $\sim1'$ northwest of the central galaxy, NGC~4696, as shown in Fig.~\ref{fig:image}. The Resolve spectrum for the field of view (FOV) is shown in Fig.~\ref{fig:spec}.
We examine Resolve spectra in the 1.8–12 keV band for the five regions shown in Fig.~\ref{fig:image}. Since the point spread function (PSF) of XRISM is relatively large ($\sim 1.3'$ for the HPD), a significant number of photons emitted from outside a given region may be seen by the telescope as "inside" the region, and vice versa.  We have accounted for this effect (referred to as spatial-spectral mixing; SSM) by fitting the spectra of the 5 regions simultaneously with plasma models and appropriate weights (Methods). We found that the spectra in each of the regions can be described by a single thermal plasma model, except for the Central region ($r\lesssim 8$~kpc from the central AGN), where two temperature components ($T=2.81$ and 1.41~keV) are required (Extended Data Table~\ref{tab:Rsolve}). For the outer regions, the temperatures are $T\sim 2.4$--3.1~keV. 

\begin{figure}[ht]
\centering
\includegraphics[width=\linewidth]{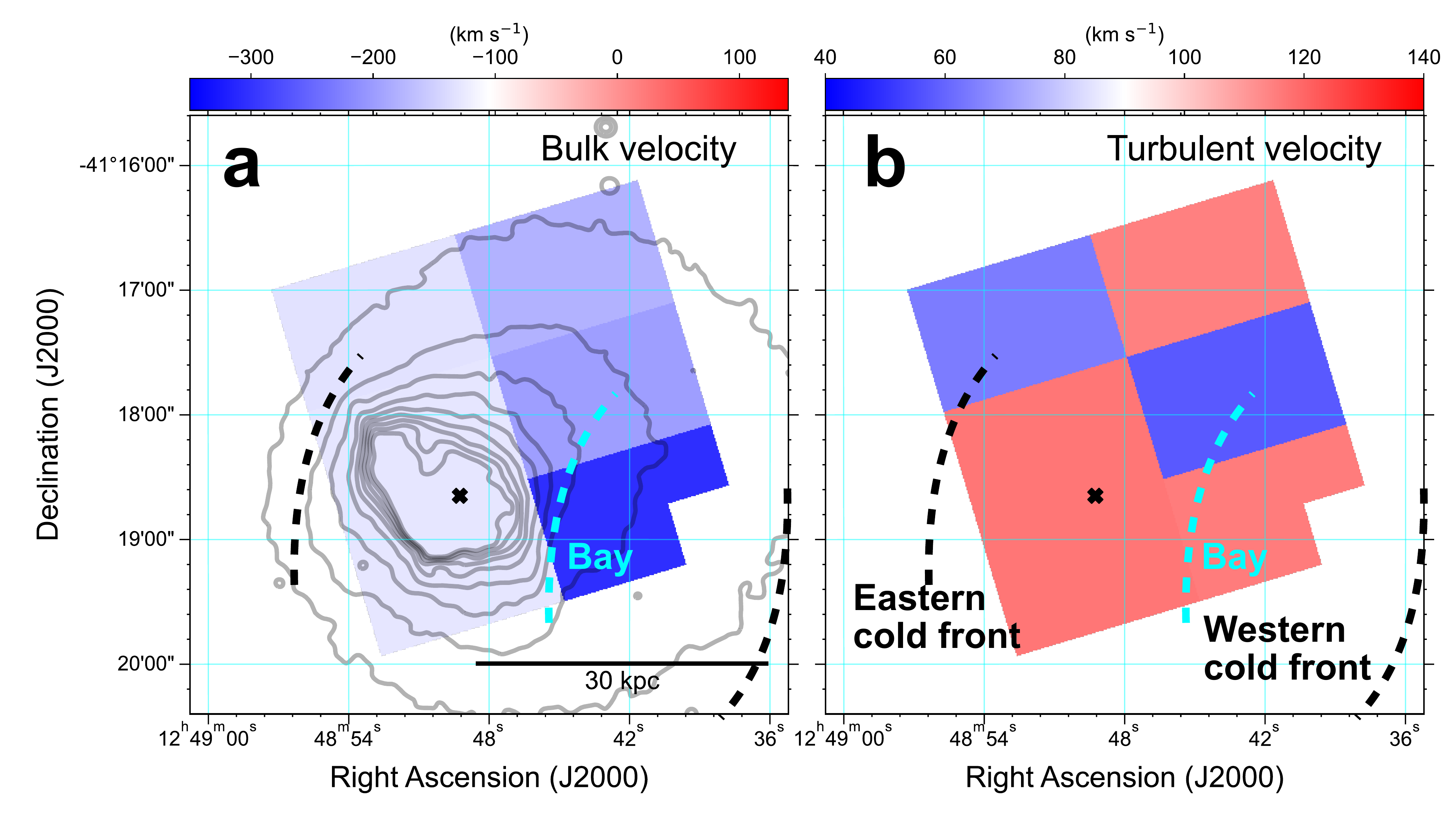}
 \caption{{\bf | Maps for {\bf a.} bulk and {\bf b.} turbulent velocities.} The positions of the AGN (marked by "X") and the "bay" are shown. The contour map of the Chandra X-ray image in 0.5--1.2 keV is overlaid on the velocity map in {\bf a.}. The black and cyan dashed lines show the cold fronts and the bay, respectively, as in Fig. \ref{fig:image}.}\label{fig:map}
\end{figure}

Using stellar absorption data from the Multi-Unit Spectroscopic Explorer (MUSE) at the Very Large Telescope (VLT), we determine that the heliocentric velocity and redshift of the stellar system of NGC~4696 are $V_{\rm N4696}=3008\pm 7\rm\: km\: s^{-1}$ and $z_{\rm N4696}=0.01003$, respectively (Methods). 
We find that the heliocentric redshifts of the ICM emission obtained with Resolve are $z\lesssim 0.0095$; they are all significantly blueshifted relative to NGC~4696, indicating that the ICM is streaming towards us relative to the galaxy, which we call the "bulk flow". In the Central region, the bulk velocity relative to NGC~4696, estimated from the redshift, is $v_{\rm bulk}= -128\pm 6\:\rm km\: s^{-1}$ (Extended Data Table~\ref{tab:Rsolve}). The bulk velocity of the ICM varies spatially, showing the presence of a velocity gradient (Fig.~\ref{fig:map}a). In the NW and W regions, the bulk velocity is $v_{\rm bulk} = -179\pm 21 \rm\: km\: s^{-1}$ and $-197^{+16}_{-17} \rm\: km\: s^{-1}$, respectively, and decreases to $-305\pm 17\rm\: km\: s^{-1}$ in the SW region. 
The Mach number of the bulk flow is $\mathcal{M}= 0.15\pm 0.01$ for the hotter component and $\mathcal{M}= 0.21\pm 0.02$ for the cooler component in the Central region and $\mathcal{M}= 0.38\pm 0.02$ in the SW region.
Previous XMM-Newton observations with the pn-CCD showed a hint of blueshifted motion near the cluster centre, although the velocity uncertainties are large \cite{2022MNRAS.513.1932G}.

The emission lines of the metals that have contaminated the gas are also broadened, indicating the presence of turbulent motion. In our 5 studied regions, the velocity dispersion $\sigma_v$ along the line of sight is systematically found to be less than $120\rm\: km\: s^{-1}$ (Fig.~\ref{fig:map}b; Extended Data Table~\ref{tab:Rsolve}), corresponding to a three-dimensional (3D) Mach number of $\mathcal{M}\lesssim0.25$ for $T\sim 2.5$~keV. The ratio of kinetic to thermal energy density is $\sim 0.03$. 
The velocity dispersion is close to that estimated with the Chandra satellite from surface brightness variations ($\sim 150\rm\: km\: s^{-1}$ on $\sim 35$~kpc scale at $r\sim 23$--50~kpc)\cite{2018ApJ...865...53Z}. In terms of 3D Mach number, the velocity dispersion in the Centaurus cluster is about half the mean value obtained for a sample of nearby giant elliptical galaxies using resonance scattering measured with the XMM-Newton Reflection Grating Spectrometer (RGS) \cite{ogorzalek2017}.
The velocity dispersion is also smaller than that observed in the Central region of the Perseus cluster with Hitomi ($163\pm 10\rm\: km\: s^{-1}$) \cite{2016Natur.535..117H,2018PASJ...70....9H}, and it does not increase significantly even in the Central region or in the AGN neighbourhood (Fig.~\ref{fig:map}b). This may indicate that the influence of the AGN on the ICM motion in the core is limited.

The increase in the absolute value of the bulk velocity $|v_{\rm bulk}|$ towards the southwest suggests that the source of the bulk flow is not the AGN at the centre of NGC~4696. The sloshing of the core gas with respect to NGC~4696 is a promising source of the bulk flow, since the motion induces the observed relative gas flow between the galaxy and the ICM \cite{2001ApJ...562L.153M,2010ApJ...717..908Z} when the direction of the oscillatory motion is nearly along the line of sight. 
The Centaurus cluster consists of the main component, including NGC 4696, called Cen~30, and a subgroup associated with NGC~4709 called Cen~45 \cite{1986MNRAS.221..453L}. 
This suggests that the cluster has not been relaxed and that internal gas motion, including sloshing, is excited in the relaxation process.
The sloshing or bulk flow in the core could be induced by the oscillation of NGC~4696 around the bottom of the cluster gravitational potential \cite{2004ApJ...610L..81H,2006ApJ...650..102A} as well as cosmological gas flows, including the large-scale velocity field produced by past cluster mergers, infalling galaxy groups and penetrating streams \cite{2017ApJ...849...54L}.

\begin{figure}[ht]
\centering
\includegraphics[width=\linewidth]{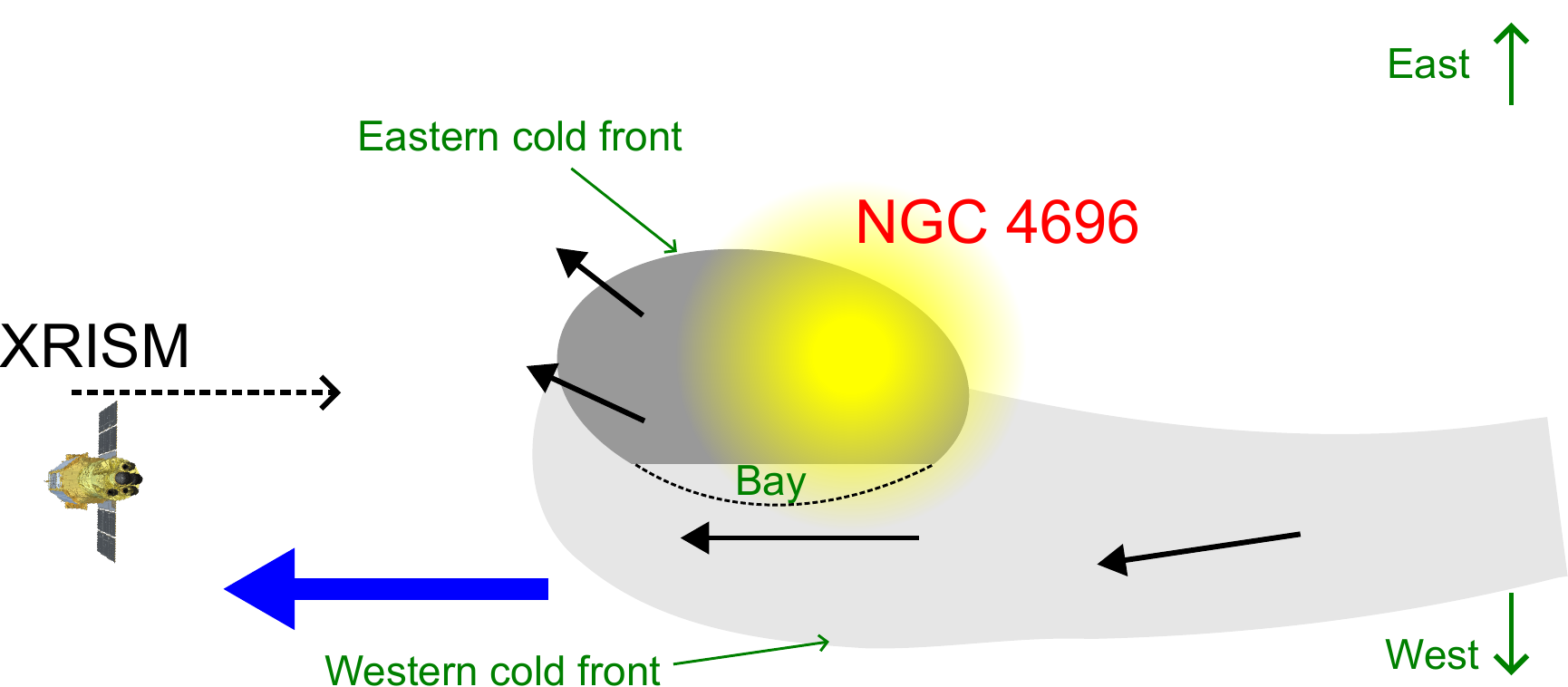}
\caption{{\bf | Schematic of the sloshing.} An east-west cross-section of the cluster core is shown. The stellar system of NGC~4696 is shown in yellow. Dark and light grey indicate the denser and thinner gas initially associated with the galaxy. The movement of the gas is indicated by the thick solid blue and thin solid black arrows. XRISM observes the system from the left.}\label{fig:cartoon}
\end{figure}

The heliocentric velocity and redshift of Cen~30 are $V_{\rm Cen30}=3118\pm 30\rm\: km\: s^{-1}$ and $z_{\rm Cen30}=0. 01040$ \cite{2024arXiv240404909V}, which are not much different from those of NGC~4696 ($V_{\rm N4696}=3008\pm 7\rm\: km\: s^{-1}$ and $z_{\rm N4696}=0. 01003$); NGC~4696 is only slightly blueshifted by $\sim 100 \rm\: km\: s^{-1}$ relative to Cen~30. 
Thus, the sloshing is probably caused by the large-scale gas motions in the cluster caused by past cluster mergers (Fig.~\ref{fig:cartoon}).
Unlike the Perseus cluster, the Centaurus cluster does not show a prominent spiral density pattern in the X-ray image, which is produced when the gas oscillates with a small orbital angular momentum \cite{2006ApJ...650..102A}. 
On the other hand, Fig.~\ref{fig:image} shows that the cluster has a pair of density jumps (eastern and western cold fronts), suggesting that the line of sight is in the orbital plane of the gas motion and we see the spiral pattern edge-on (Fig.~\ref{fig:cartoon}). The density jumps correspond to the surface of the spiral arms (see also Fig.~19 in Ref~\cite{2006ApJ...650..102A}). 
Fig.~\ref{fig:map}a shows an east-west gradient in the bulk velocity, which may reflect the rotational motion associated with the spiral pattern. 

The bulk velocity is particularly large in the circular structure called the ``bay'' in the Chandra image (Figs.~\ref{fig:image} and~\ref{fig:map}a). Previous studies have suggested that the bay may be a Kelvin-Helmholtz instability roll \cite{2017MNRAS.468.2506W}.
However, if this is the case, there should be a strong gas flow along the bay in the plane of the sky, which is not clear from our observations. Alternatively, the bay could have been formed by a locally strong flow along the line of sight that washed away an old cavity created by AGN activity in the past. 

The bulk flow could help to suppress the development of massive cooling flows by shifting low entropy gas away from the centre of NGC 4696, Indeed, Chandra observations have shown that the X-ray peak appears to be  shifted $\sim 1$~kpc away from the central AGN (Extended Data Fig.~\ref{fig:cavity}). This has been observed in other clusters as well \cite{2016MNRAS.460.2752W}.
The gradient of the bulk velocity and the acceleration and deceleration of the core gas associated with the sloshing could excite Kelvin-Helmholtz and Rayleigh-Taylor instabilities, eventually leading to the formation of large eddies. The eddies could mix the gas and help prevent excessive cooling in the galaxy by transferring thermal energy from the surrounding hot ICM into the galaxy \cite{2004ApJ...612L...9F,2010ApJ...717..908Z}. The eddies could also distribute the locally injected energy from the central AGN and contribute to stable heating. Part of the kinetic energy of the bulk flow could also be converted into thermal energy by the formation and decay of eddies. 
From the density distribution (Extended Data Fig. ~\ref{fig:nTZ}), the gas mass in the Resolve FOV is estimated to be $\sim 10^{10}\: M_\odot$. For a bulk velocity of $|v_{\rm bulk}|\sim 200\rm\: km\: s^{-1}$, the available kinetic energy is $\sim 5\times 10^{57}$~erg.
This is comparable to the energy deposited in cavities ($\sim 6\times 10^{57}$~erg; Extended Data Table \ref{tab:cavity_powers}).
We note that the typical duration of the bulk flow cannot be too long since, otherwise, the metal abundance peak observed in the centre of NGC~4696 should have been blown away.
Since the scale of the peak is $R_{\rm p}\sim 20$~kpc (Extended Data Fig.~\ref{fig:nTZ}) and the bulk flow speed is $|v_{\rm bulk}|\sim 200\rm\: km\: s^{-1}$, the duration should not be $\gg R_{\rm p}/|v_{\rm bulk}|\sim 10^8$~yr.
We also note that the decrease in abundance at the cluster centre derived from the Chandra spectrum (Extended Data Fig.~\ref{fig:nTZ}) could be caused by something other than the bulk flow (e.g., dust formation for the cooler gas \cite{2016MNRAS.457...82S}).

In contrast to the X-ray gas, the cold ($\lesssim 100$~K) and warm ($\sim 10^4$~K) gases detected in the central few kpc of NGC~4696 do not show the bulk motion relative to the galaxy \cite{2011MNRAS.418.2386M,2011MNRAS.417.3080C,2019A&A...631A..22O,2023PASJ...75..925F}, which is different from some other clusters \cite{2024arXiv240402212G}. 
Rather, they appear to rotate around the AGN in nearly the north-south direction at a speed of $\sim$ 250--300 km s$^{-1}$ (Extended Data Fig.~\ref{fig:muse}). Their unique motion, unrelated to the X-ray gas bulk flow, suggests that these denser gases may have cooled from the hot X-ray gas and accumulated in the galactic centre before the bulk flow began to blow. The rotational velocity of $\sim 300\:\rm km\: s^{-1}$ is probably much greater than the hot gas originally had since the velocity increases as the cooling gas moves inward due to the conservation of angular momentum.
Because the cold and warm phases are confined to the bottom of the galactic potential well, they are relatively immune to the bulk flow. However, the outer part of the cold and warm gases can be blown away by the bulk flow, preventing too much of it from being deposited in the galactic centre. 
The rotation of the cold and warm components is unlikely to have much effect on the bulk flow motion of the X-ray gas because of their small covering fraction. 

The dissipation of turbulence is expected to heat the gas \cite{2014Natur.515...85Z}.  
The turbulent heating rate in the gas with mass density $\rho$ is $Q_{\rm turb} \sim 5\: \rho \sigma_v^3/l_t$, where $l_t$ is the length scale \cite{2014Natur.515...85Z}. 
The radiative cooling rate of the gas is calculated from the gas density and the mean temperature $T$: $Q_{\rm cool}=n_e n_i \Lambda_n(T)$, where $n_e$ and $n_i$ are the electron and ion number densities respectively, and $\Lambda_n(T)$ is the normalised cooling function \cite{1993ApJS...88..253S}.
For the Central region, $\sigma_v\sim 117\rm\: km\: s^{-1}$, $l_t\sim 10$--20~kpc (the size of the bright region at the cluster centre or that of the Central region in Fig. ~\ref{fig:image}), the mean temperature $T\sim 2$~keV and the metal abundance $Z\sim 2$~solar (Extended Data Table~\ref{tab:Rsolve}), while $\rho$, $n_e$ and $n_i$ are obtained from Chandra observations ($n_e\sim 0.02\rm\: cm^{-3}$; Extended Data Fig.~\ref{fig:nTZ}). From these values we have $Q_{\rm turb}/Q_{\rm cool}\sim 0.5$--1.1, suggesting that the turbulent heating rate may be comparable to the radiative cooling rate, although this $Q_{\rm turb}$ estimate is only approximate. 
The main source of turbulence appears to be sloshing rather than past AGN activity, since there is no significant increase in turbulent velocity around the AGN (Fig.~\ref{fig:map}b). If this is the case, then the AGN contribution to $Q_{\rm turb}$ is limited.
The high central Fe abundance value (Extended Data Fig. ~\ref{fig:nTZ}) indicates a very long metal accumulation time \cite{2004A&A...416L..21B} and little AGN-induced perturbation. This is consistent with our conclusion that despite the presence of several generations of AGN-induced cavities indicating jet powers reaching $\sim 3\times10^{43}$ erg s$^{-1}$ (Extended Data Table \ref{tab:cavity_powers}), which is comparable to the radiative cooling rate of the core ($\sim 3\times10^{43}$ erg s$^{-1}$) \cite{2004ApJ...607..800B}, the kinematic influence of central AGN in the Centaurus cluster through turbulence is not significant.
For comparison, in the Perseus cluster, where the cavities were found to be associated with regions of increased turbulence \citep{2018PASJ...70....9H}, the estimated jet powers and total energies are at least an order of magnitude larger \citep[e.g.][]{rafferty2006}.

Our results suggest that, at least on short timescales ($\sim 10^8$~yr), the bulk gas motion induced by sloshing plays a more important role in the thermal balance in the cluster core than the smaller-scale motion induced by AGN cavities (Extend Data Fig.~\ref{fig:cavity}), which is observed by Resolve as line broadening or turbulence.

\clearpage



\clearpage

\clearpage

\section{Methods}\label{sec11}

\subsection{XRISM Data reduction and analysis}

The Resolve instrument onboard XRISM \citeappx{Ishisaki2022} is a system that combines an X-ray microcalorimeter spectrometer with an X-ray Mirror Assembly (XMA) to cover a $3' \times 3'$ field of view (FOV). 
The microcalorimeter array is operated in a dewar in which a multi-stage cooling system maintains a stable environment at 50 mK. 
The Resolve instrument was originally designed to cover the energy range between 0.3 and 12 keV, but only data in the $E\gtrsim 2$ keV band were available during the Centaurus observation, because the Resolve dewar gate valve, which consists of a Be window and its support structure, was still closed at the time of the observation. In this paper, we use only the Resolve data to study the velocity structure of the Centaurus cluster core.

The Centaurus cluster observational data were processed using pre-pipeline software version "004\_001.15Oct2023\_Build7.011" and the processing (pipeline) script "03.00.011.008". In the following data analysis, the CALDB8 and xa\_rsl\_rmfparam\_20190101v006.fits were used to make the redistribution matrix files (RMFs) as is standard practice. The gain correction of Resolve was performed with $^{55}$Fe radioactive sources on the filter wheel (FW) and a total of 58 gain fiducial checks were performed during the observation, for $\sim$30 minutes once per orbit. The full width at half maximum (FWHM) of the composite Mn K$\alpha$ spectrum of all pixels except for the calibration pixel (pixel 12) for the fiducial intervals was $4.50\pm0.01$ eV. 
The FWHM energy resolution as a function of energy is given per pixel in the corresponding CALDB file, which is based on the initial analysis of PV phase calibration data. The uncertainty on the FWHM energy resolutions is estimated to be about 0.3 eV. The CALDB values of per-pixel FWHM energy resolution were used in the calculation of RMFs used for spectral analysis.

The energy scale accuracy after gain reconstruction has been shown to be 0.2 eV or better in-flight for a flat-field average in the band 5.4--8.0 keV using onboard calibration sources \citeappx{Porter2024,Eckart2024}. Calibration measurements of the broad-band energy scale from 2 to 9 keV, in-flight, are consistent with an error of  $\lesssim$1~eV for a flat-field average. The Resolve instrument includes a calibration pixel that is part of the main detector but located just outside the instrument aperture that is constantly illuminated with a collimated $^{55}$Fe radioactive source. The calibration pixel can be used to assess the energy-scale reconstruction outside of the gain fiducial measurements by reconstructing its gain only during the gain fiducial intervals using the same method as the main array, but then fitting the calibration line only during the on-target intervals. For this observation, the calibration pixel gave an energy scale error of $0.195\pm0.05$ eV at 5.9 keV, which is comparable to other XRISM observations. The gain scale error of 0.2 eV at 6 keV corresponds to a gas motion of $\sim$10 km s$^{-1}$, which is comparable to the statistical errors of bulk motion for the Central region and smaller than those for the other regions in our measurement. On the other hand, the error of about 0.3 eV in the line spread function (LSF) corresponds to the systematic uncertainty on the turbulent velocity of $\sim 6$ km s$^{-1}$ for 100 km s$^{-1}$ if He-like Fe lines drive the spectral fits as a major factor of our analysis. 
If the spectral fits are determined by lower energy emission lines around 2.5 keV such as sulfur lines, the systematic uncertainties are quite a bit higher ($\sim$35 km s$^{-1}$ uncertainty for turbulent velocity of 100 km s$^{-1}$), although a minor effect in our analysis.
In fact, when we performed the spatial-spectral mixing (SSM) fits using only the spectra above 5 keV, the resultant values of turbulent velocities and redshifts for all regions did not change within the statistical errors from the results of the original broadband (1.8--12~keV) fit. Here only the lower temperature component in the Central region in the >5 keV fit was fixed to the value in the broadband fit. 

Resolve events are classified as high-, medium-, and low-resolution (H, M, and L) grades based on the time proximity to a successive event in the same pixel, and further, they are subdivided into primary and secondary (p and s) classes depending on whether there is a triggered event after or before the event within a specific time interval \citeappx{Ishisaki2018}. The performance quoted here is for High-primary (Hp) events, the grade that was used in our analysis and which represents $>$98\% of the counts from this source. Finally, pixel 27, which is located in the “N” region in Fig. \ref{fig:image}, has been shown to exhibit gain jumps of a few eV on time scales of a few hours, and thus is not currently recommended to be included in Resolve observations. However, in this early observation, gain fiducial measurements were made every 90 minutes, and the gain jumps were corrected. We, thus, used the full array for the spectral analysis of the entire observation. 

\begin{figure}[ht]
\centering
\includegraphics[width=\linewidth]{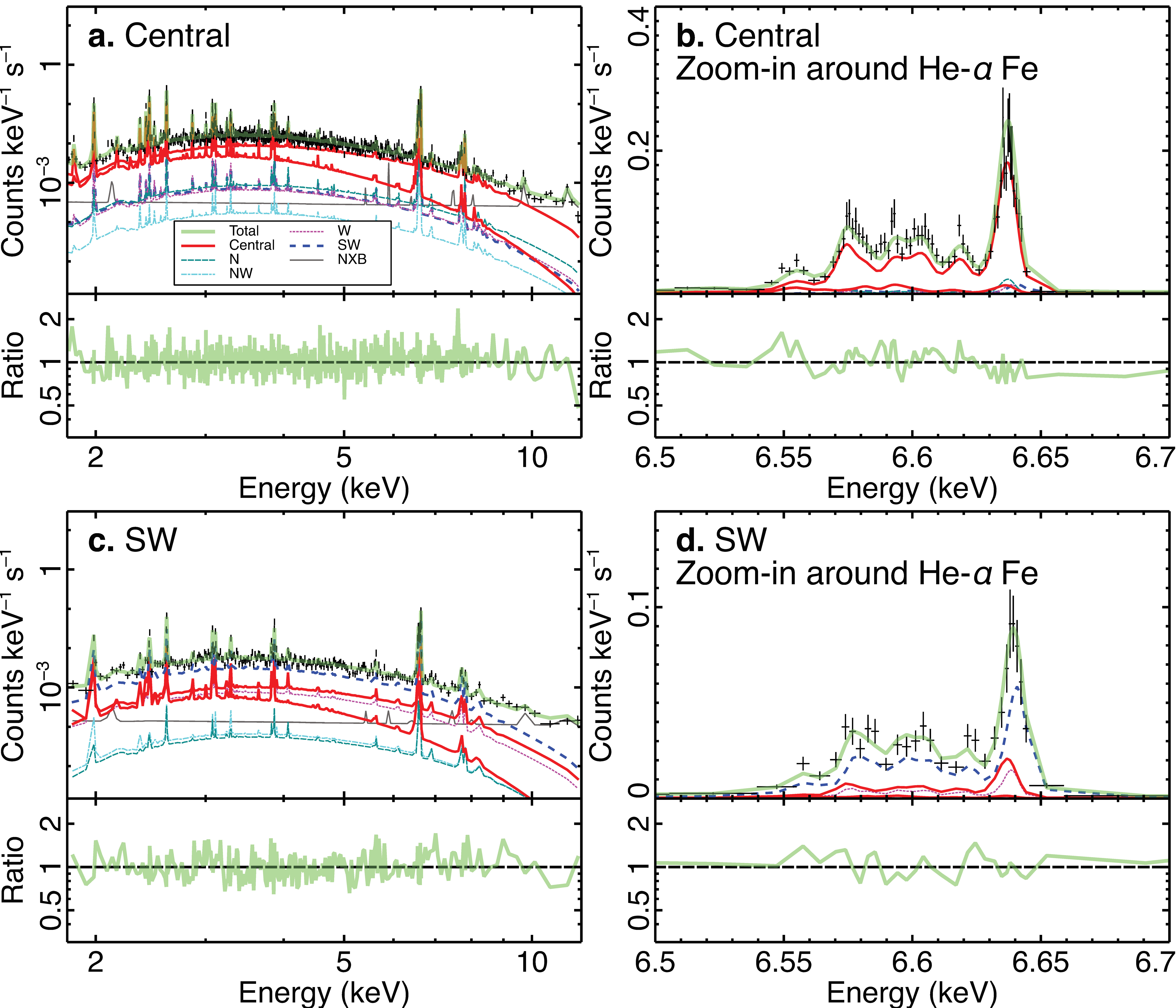}
\caption{{\bf | Representative spectra from and the best-fitting models, including the SSM effect.} {\bf a.} Broadband and {\bf b.} Zoom-in spectra around He-$\alpha$-Fe lines for the Central region. {\bf c.} and {\bf d.} are those for the SW region.
Green thick solid line in upper panels show the sum of the total emissions from all the regions. The grey thin solid lines indicate the NXB model. Each coloured line except for the green represents the contribution from one region to the total spectrum in the centre (left) and SW (right) regions. Note that two temperature components in the Central region are shown in two red thick solid lines.}\label{fig:ssmfit}
\end{figure}

\begin{figure}[ht]
\centering
\includegraphics[width=\linewidth]{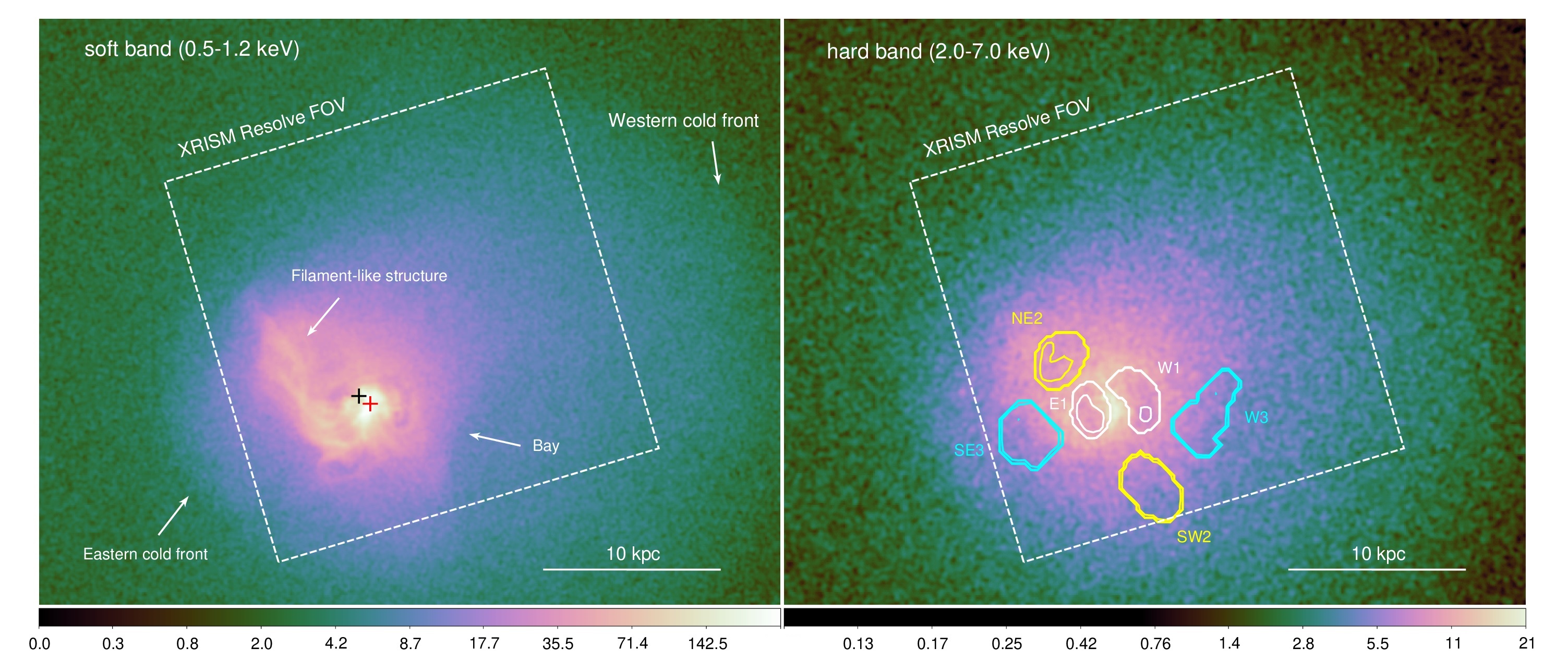}
\caption{{\bf | Chandra X-ray images of the Centaurus core.}
\textbf{Left}: Soft band ($0.5$--$1.2\:$keV) image with main features highlighted. The white dashed line shows the XRISM Resolve FOV. The black cross indicates the position of the central AGN of galaxy NGC4696 separated by $\sim 1\,$kpc from the X-ray peak of the Centaurus cluster (red cross). \textbf{Right:} Hard band ($2.0$--$7.0\:$keV) image with  the best estimates of the cavity borders by CADET for individual detected cavity generations.
}\label{fig:cavity}
\end{figure}

\begin{figure}[ht]
\centering
\includegraphics[width=0.97\linewidth]{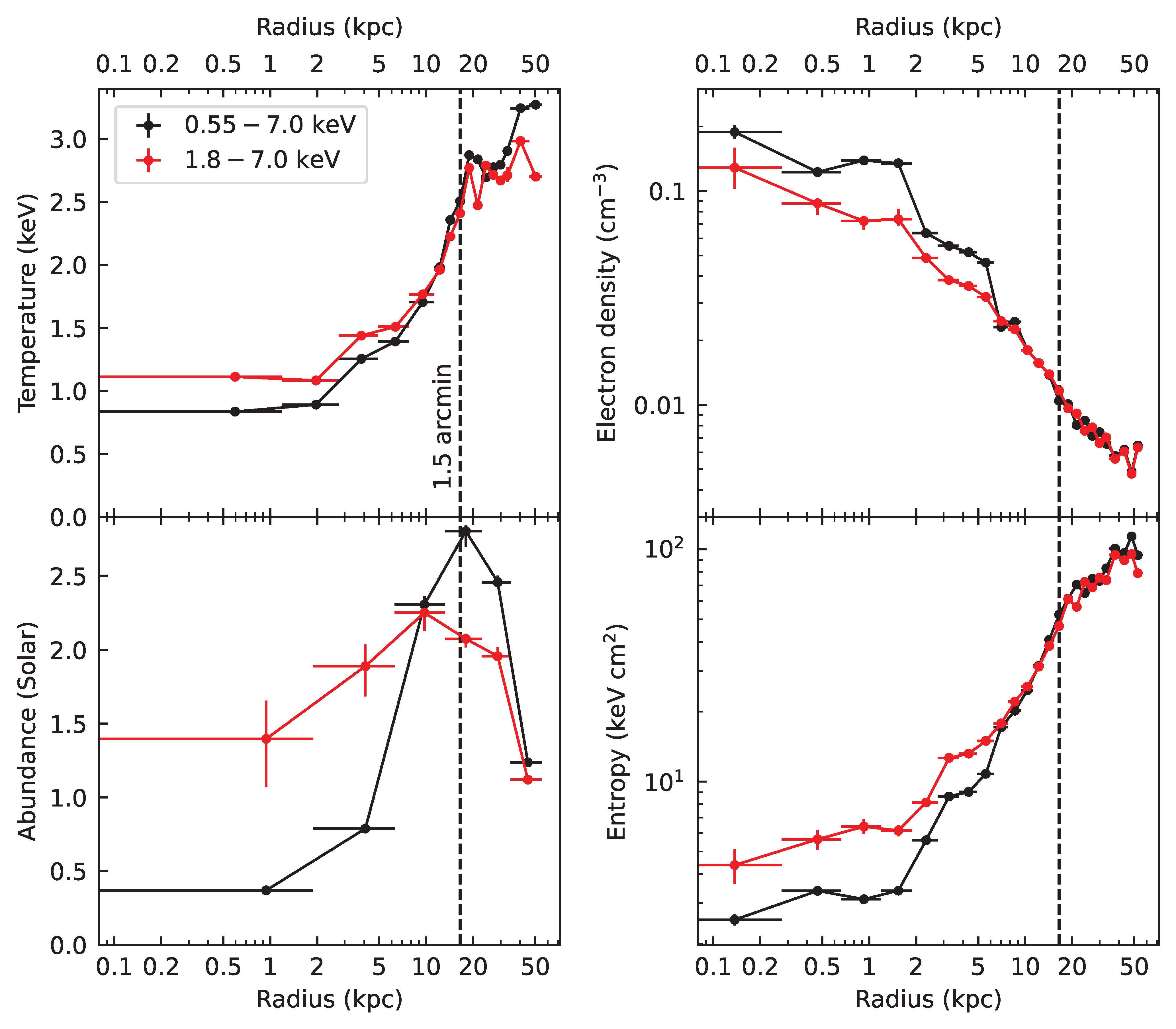}
\caption{{\bf | Temperature, electron density, metal abundance, and entropy profiles.} They are obtained from deprojection analysis of broad (black points) and hard (red points) band Chandra data.}\label{fig:nTZ}
\end{figure}

We applied the standard event screening criteria \citeappx{Kilbourne2018, XRISM2024b}
for the Centaurus cluster observations, discarding time intervals of eclipses by Earth, small elevation angles from the sunlit Earth limb ($< 20^\circ$), passing through the South Atlantic Anomaly, and the cryo-system recycling. This paper used only the Hp detector events (ITYPE~$=0$) for the spectral analysis. 
The non-X-ray background (NXB) spectra are estimated from the stacked Night Earth database (NXBDB) using the \texttt{rslnxbgen} task; then we applied the same screening as above, except for the threshold parameter of the dark Earth. We fitted the NXB spectra with an updated version of the Resolve NXB model reported in \citeappx{XRISM2024b}, and confirmed the instrumental line intensities. In the spectral fits, the NXB components were included as the model. Note that the Mn K$\alpha$ line scattered from the collimated $^{55}$Fe source for the calibration pixel is non-uniformly distributed across the array, and care should be taken in treating subarray regions until more NXBDB statistics are available. However, the NXB contribution has a small impact on determining the central energy of emission lines and the line width from our spectral analysis since the positions of the instrumental lines are away from the emission lines in the source spectrum on energy space as shown in Extended Data Fig. \ref{fig:ssmfit}.
We created the RMFs using the extra-large mode option in the \texttt{rslmkrmf} task to represent the spectral shape in the lower and higher energy bands, including the escape peak and electron loss continuum effects. We also created ancillary responses (ARFs) for the spectrum of each region. The ARFs were appropriate for the diffuse emission from the ICM and were based on the X-ray image observed with Chandra in the 2--8 keV energy band using the \texttt{xaarfgen} task. Finally, the spectra were grouped so that each bin had at least 1 count to fit the spectra using the C-statistic [\citeappx{1979ApJ...228..939C}] without subtracting any background emissions.

Here, the ICM emissions in each region were reproduced by one or two thermal plasma models (bvapec model with AtomDB version 3.0.9 in the XSPEC package).
First, we adopted a single (1T) and double (2T) temperature model in the spectral fits for each region. As a result, only the spectrum of the Central region needed the 2T model, and those in the other regions were well represented by the 1T model on the statistics. The abundances of Si, S, Ar, Ca, Mn, Cr, Fe and Ni were freely variable. The He abundance is fixed at 1 solar. The abundances for other elements were fixed to be 1.5 solar. In the 2T model for the Central region, the parameters between hotter and cooler components except for the temperatures and normalizations were linked for each other.
The HPD of Resolve XMA is 1.3$'$, thus contamination from adjacent regions must be considered. This effect is taken into account in the SSM analysis, and the amount of leakage into each region is evaluated by the ARFs of each region produced using ray-tracing, taking into account the shape of the Point Spread Function (PSF). We note that the X-ray spectrum in a region includes photons from neighbouring regions due to the PSF of the XMA and the effect of small wobbles in attitude. 
We calculated the SSM effect, which takes into account the flux from one region of the astronomical model and distributes it over several observation regions (\cite{2018PASJ...70....9H} for more details).
According to ray-tracing calculations, the leakage from the Central region to the SW region is equivalent to $\sim 20$\% of the SW region, and about half of the total flux in the SW region would come from the other regions. Therefore, it is taken into account in the SSM analysis. The velocity difference between the centre and SW regions is $\sim 110$ km s$^{-1}$ when the spectra for each region were fitted independently, without the SSM analysis, therefore, the SSM analysis emphasizes the velocity difference more. We provide representative spectra and best-fitting models with the SSM analysis in Extended Data Fig.~\ref{fig:ssmfit}. The derived parameters are summarised in Extended Data Table~\ref{tab:Rsolve}. 
We have confirmed that there are no strong correlations between bulk and turbulent velocities and temperature. Note that the resultant ratio of the emission measure of the hotter to cooler component was 0.94$\pm$0.26\@. If the turbulent velocities of the two temperature components in the centre were not tied, we did not see the significant difference within their systematic errors.
As for the systematic error of this SSM analysis, Ref~\citeappx{Hayashi2024} reports that when the point-like source was on-axis, 
the difference between observed count rates and those predicted by ground data was less than 20\% for most pixels (although much larger for some pixels)
and when it was 1.8$'$ off-axis, the difference was less than 50\% for each pixel \citeappx{Hayashi2024, Tamura2024}.  
The differences between observed count rates per pixel and those predicted by the XMA model in the CALDB are somewhat larger than the differences between inflight and ground data. 
In our case, since the Centaurus cluster is a diffuse source, the effect is difficult to evaluate, but we have investigated the impact of the PSF contamination by artificially shifting the input Chandra image for the ARF calculations by 10$''$ in the on-axis or off-axis direction to investigate the systematic error. Since the difference between the results obtained here and those in  Extended Data Table~\ref{tab:Rsolve} was less than the statistical errors, we evaluated these systematic errors to be less than the statistical errors, and it does not affect our results.
All the abundances are expressed using the proto-Solar values \citeappx{2009LanB...4B..712L}. The foreground absorption column was fixed at $7.77\times 10^{20}\rm\: cm^{-2}$ \citeappx{2016A&A...594A.116H}. We have ensured that the X-ray emission from the central AGN does not affect our results by using the 3$\sigma$ upper limit of $1.2 \times 10^{40}$ erg s$^{-1}$ in 2--10 keV band with a photon index of 1.7 \citeappx{2006MNRAS.365..705T}. We adopted the heliocentric correction of $24\rm\: km\: s^{-1}$.

\subsection{Redshift determination}

\begin{figure}[ht]
\centering
\includegraphics[width=0.97\linewidth]{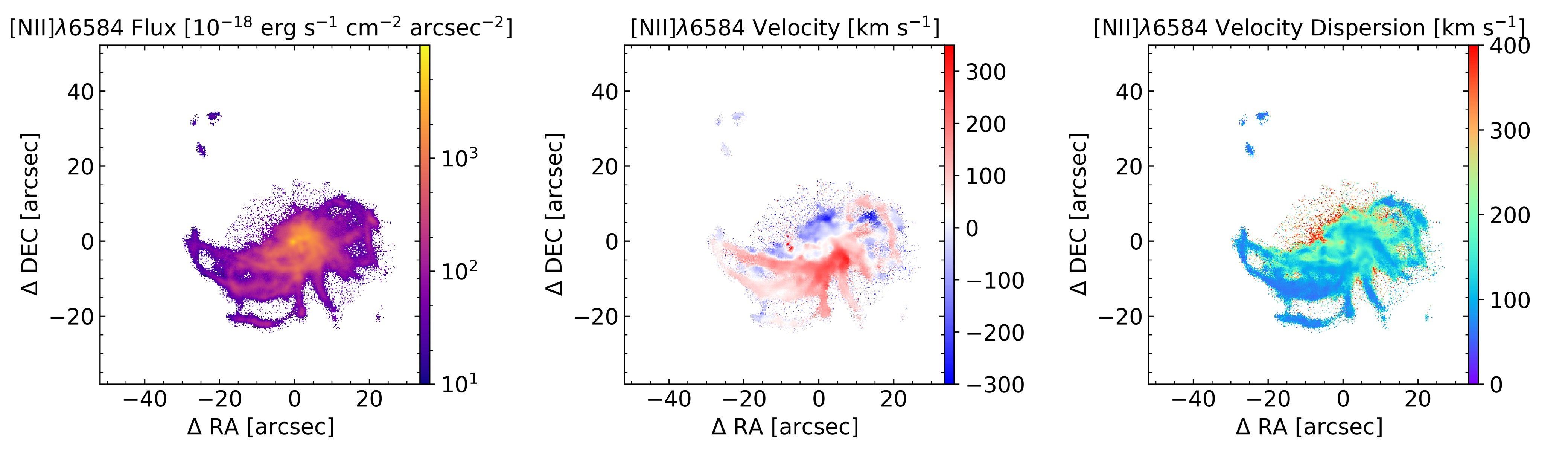}
\caption{{\bf | Flux, velocity and velocity dispersion maps of the warm, ionised gas at the centre of NGC~4696.} They are traced by the strongest optical line \NII{} $\lambda$6584 from the {\em MUSE} data. The nucleus is at the (0, 0) position. The velocity is relative to the system velocity of 3008 km/s.}
\label{fig:muse}
\end{figure}

The Centaurus cluster has two main subclusters, Cen 30 and Cen 45 \cite{2024arXiv240404909V}. NGC~4696 is the central galaxy of the Cen 30 subcluster. The redshift of the Cen 30 subcluster has been constrained to $z$=0.0104, or the velocity of 3118$\pm 30 \rm\: km\: s^{-1}$\cite{2024arXiv240404909V}. We also used the Multi-Unit Spectroscopic Explorer (MUSE) data of NGC~4696 to determine its velocity from stellar absorption lines. Twenty-two MUSE observations from 2014 to 2023 were used, with a total exposure time of 5.66 hours, under seeing conditions of 0.70$''$--1.21$''$ and airmass of 1.08--1.72. For each pointing, we used the MUSE pipeline (version 2.9.0) with the  ESO Recipe Execution Tool (EsoRex) to reduce the raw data, which provides a standard procedure to calibrate the individual exposures and combine them into a datacube. Further sky subtraction was performed with the Zurich Atmosphere Purge software (ZAP).
We extracted the MUSE spectrum within the central 1 kpc radius and fit it with pPXF \citeappx{2004PASP..116..138C}. The derived heliocentric stellar velocity is 3008$\pm7\rm\: km\: s^{-1}$. We also examined the spectrum within the central 4 kpc radius, with the same stellar velocity derived.
With the pPXF fits to the MUSE data, we also derived the emission line flux, velocity and velocity dispersion maps for the optical emission-line nebula. Stellar absorption has been corrected before emission-line fits. The maps for the brightest emission line \NII{} 6584 \AA{} are shown in Extended Data Fig.~\ref{fig:muse}. 

\subsection{Complementary Chandra observations}

We leveraged the great spatial resolution of the Chandra X-ray observatory to perform a complementary spatially-resolved image and spectral analysis of the Centaurus cluster. For the analysis, we used 15 archival Chandra observations with a total cleaned exposure time of 777.4\:ks. All Chandra OBSIDs were processed using the \texttt{chandra\_repro} script (CalDB 4.11.3), reprojected, and merged. The point sources were detected using the \texttt{wavdetect} procedure (ECF$\:=0.9$, threshold$\:=10^{-5}$), and the background light curves in the $0.5$--$7.0\:$keV band were extracted and deflared using the \texttt{deflare} routine ($3\sigma$ clipping level).

For the spatially resolved spectral analysis, we limited our analysis to only 13 observations with the ACIS-S3 chip centred at the central galaxy, NGC\,4696. Using the \texttt{specextract} procedure, we produced spectra for 24 concentric annuli centred at the position of the AGN with the outermost annulus of 300 arcsec, where background spectral files for each annulus were extracted from blank-sky backgrounds. The spectral modelling was performed using PyXspec v2.1.3 package (XSPEC 12.13.1 \citeappx{1996ASPC..101...17A}). The thermodynamic profiles were obtained by fitting a single temperature \texttt{apec} model absorbed by the photoelectric absorption (\texttt{phabs}). The deprojection was performed using the \texttt{projct} mixing model component, assuming spherical symmetry. In addition, we included a power-law model component with a photon index of 1.9 \citeappx{2007A&A...463...79G} to describe the emission of the central AGN in the innermost annulus and in outer annuli with a photon index of $\Gamma = 1.56$ \citeappx{2003ApJ...587..356I} to account for unresolved point sources. The model for each annulus is identical and is defined as follows:\\
\begin{center}
    \texttt{projct(phabs(apec)) + phabs(powerlaw)}
\end{center} \vspace{2mm}
The temperatures and metal abundances of individual annuli were allowed to vary. However, for some of the annuli, we tied parameters for 2 or more neighbouring annuli together (see Extended Data Fig. \ref{fig:nTZ}) to constrain these parameters with a relative uncertainty smaller than 25 percent. The fitting was performed in the broad energy band ($0.55$--$7.0\:$keV) as well as only in the hard band ($1.8$--$7.0\:$keV), which is more similar to XRISM Resolve spectra. During the fitting, the redshift was fixed to $z = 0.01040$, and the galactic hydrogen column was fixed to the value of $N_H=7.77\times 10^{20}\rm\: cm^{-2}$ \citeappx{2016A&A...594A.116H}, however, even after thawing the parameter we obtained consistent results. The background files were treated using the \texttt{wstat} statistics (modified \texttt{cstat}; \citeappx{1979ApJ...228..939C}). For the analysis of the Chandra data, we assumed the standard flat $\Lambda$CDM cosmology, with $H_0 = 70$\:km s$^{-1}$ Mpc$^{-1}$, $q_0 = 0$, $\Omega_{\Lambda,0} = 0.73$. All the abundances are expressed with respect to proto-Solar values \citeappx{2009LanB...4B..712L}. The uncertainties are expressed in the 1$\sigma$ credible interval and were estimated from posterior distributions obtained from MCMC simulations. The resultant profiles of temperature, electron density and metal abundance are shown in the Extended Data Fig. \ref{fig:nTZ}. On top of that, we also calculated the profile of specific entropy ($K = kT / n_{\text{e}}^{2/3}$) to show that the profile is flattening in the centre, which is a sign of a cool core.

Besides the spatially resolved analysis, we also performed an analysis of the Chandra spectrum extracted from the Central region (see Fig. \ref{fig:image}). The spectrum has been modelled using a 2T \texttt{vapec} model, where abundances of both \texttt{vapec} components were tied together. The fitting was performed both in the broad Chandra band ($0.5$--$7.0$ keV) as well as in the hard band with a low energy limit comparable to that of the XRISM Resolve instrument ($1.8$--$7.0$ keV). For the broadband fit, the temperatures of the individual \texttt{vapec} components were estimated to be $0.96\pm0.03$\:keV and $1.87\pm0.04$\:keV, and for the hard-band fit, the temperatures of the individual components were $1.26\pm0.06$\:keV and $2.47\pm0.15$\:keV, which are slightly lower compared to the temperatures estimated from the XRISM Resolve spectrum for the Central region (Extended Data Table~\ref{tab:Rsolve}), but still consistent within statistical errors.
The reason why the broadband fit gives lower temperature values appears to be the presence of the low-temperature gas component, which emits in the soft band but does not contribute significantly to the hard band.

Furthermore, we generated the merged images in the broad energy band ($0.5$--$7.0$\:keV) as well as in individual sub-bands: soft ($0.5$--$1.2$\:keV), medium ($1.2$--$2$\:keV), and hard ($2.0$--$7$\:keV). We filled the point sources in these images using the \texttt{dmfilth} routine and exposure-corrected them with weighted exposure maps. We then used these images to identify individual spatial structures such as cold fronts, X-ray cavities, etc. Because the soft energy band is dominated by the filament-like sub-structure, X-ray cavities were detected mainly based on the hard-band image (we also tested other sub-bands, e.g. $1.5$--$7$\:keV or $2.5$--$7$\:keV, but the obtained results were consistent). For detecting the X-ray cavities, we used the Cavity Detection Tool, and the significance of the cavity detections was tested using azimuthal and radial surface brightness profiles \citeappx{10.1093/mnras/stad3371}. The CADET pipeline successfully detected the previously known central cavity pair (Generation 1). On top of that, the pipeline detected an older generation of cavities in the NE-SW direction (Generation 2) and potentially an even older but non-significant
generation of cavities in the SE-W direction (Generation 3) (Extended Data Fig. \ref{fig:cavity}). 
These cavities were also detected in an earlier study using Chandra data \cite{2016MNRAS.457...82S}.
The total energy stored in the detected X-ray cavities and the corresponding mechanical jet powers were estimated by combining the cavity extent estimated with CADET and temperature and density profiles obtained from the spatially-resolved spectral analysis of Chandra data (Extended Data Table \ref{tab:cavity_powers}). Including the energy released by the potential Generation 3 X-ray cavities, the total energy released into the hot atmosphere of NGC4696 in the last 15 Myr is approximately $\sim6\times10^{57}$ erg (or~$\sim4 \times 10^{57}$ erg in last 12 Myr in the case of 2 latest cavity generations).

\vspace{5mm}

\noindent
\textbf{Data and code availability}
The observational data analysed during this study will be available in the NASA HEASARC repository (https://heasarc.gsfc.nasa.gov/docs/xrism/) in the summer 2025. The atomic databases used in this study are also available online
(AtomDB, http://www.atomdb.org/).

\clearpage

\bibliographystyleappx{sn-mathphys-num}

\clearpage

\noindent
\textbf{Acknowledgements}
This work was supported by JSPS KAKENHI grant numbers JP22H00158, JP22H01268, JP22K03624, JP23H04899, JP21K13963, JP24K00638, JP24K17105, JP21K13958, JP21H01095, JP23K20850, JP24H00253, JP21K03615, JP24K00677, JP20K14491, JP23H00151, JP19K21884, JP20H01947, JP20KK0071, JP23K20239, JP24K00672, JP24K17104, JP24K17093, JP20K04009, JP21H04493, JP20H01946, JP23K13154, JP19K14762, JP20H05857, and JP23K03459, and NASA grant numbers 80NSSC23K0650, 80NSSC20K0733, 80NSSC18K0978, 80NSSC20K0883, 80NSSC20K0737, 80NSSC24K0678, 80NSSC18K1684, and 80NNSC22K1922. LC acknowledges support from NSF award 2205918. CD acknowledges support from STFC through grant ST/T000244/1. LG acknowledges financial support from Canadian Space Agency grant 18XARMSTMA. AT and the present research are in part supported by the Kagoshima University postdoctoral research program (KU-DREAM). SY acknowledges support by the RIKEN SPDR Program. IZ acknowledges partial support from the Alfred P. Sloan Foundation through the Sloan Research Fellowship. MS acknowledges the support by the RIKEN Pioneering Project Evolution of Matter in the Universe (r-EMU) and Rikkyo University Special Fund for Research (Rikkyo SFR). NW and TP acknowledge the financial support of the GAČR EXPRO grant No. 21-13491X. Part of this work was performed under the auspices of the U.S. Department of Energy by Lawrence Livermore National Laboratory under Contract DE-AC52-07NA27344. The material is based upon work supported by NASA under award number 80GSFC21M0002. This work was supported by the JSPS Core-to-Core Program, JPJSCCA20220002. The material is based on work supported by the Strategic Research Center of Saitama University.

\noindent
\textbf{Author Contributions}
As the leader of the Centaurus cluster target team in the XRISM Science Team (XST), Y. Fujita led this research project and wrote the manuscript. K. Sato is the sub-leader of the target team, led the data analysis, and prepared the manuscript. He also contributed to the Resolve hardware development, integration tests, launch campaign, in-orbit operation, and calibration. K. Fukushima, F. Mernier, K. Matsushita, A. Simionescu, M. Kondo and A. Majumder analysed the data. Y.F., K.S., K.F., F.M., K.M., A.S. and K. Nakazawa discussed the results. T. Hayashi, T. Okajima, K. Tamura, Y. Maeda, M. Lowenstein, T. Yaqoob, E. Miller, M. Markevitch, F. Porter, M. Leutenegger, C. Kilbourne, and R. Kelley contributed useful comments on data calibration and systematic errors, as well as on the content of the manuscript.  M. Sun, K. Hosogi, N. Werner and T. Pl\v{s}ek provided additional data and analysis. R. Mushotzky, I. Zhuravleva, and E. Behar helped to improve the manuscript. 
The science goals of XRISM were discussed and developed over 7 years by the XST, all members of which are authors of this manuscript. All the instruments were prepared by the joint efforts of the team. The manuscript was subject to an internal collaboration-wide review process. All authors reviewed and approved the final version of the manuscript.

\noindent
\textbf{Author Information}
Reprints and permissions information is available at www.nature.com/reprints. The authors declare no competing financial interests. 
Correspondence and requests for materials should be addressed to Y.F. (y-fujita@tmu.ac.jp) and K.S. (ksksato@post.kek.jp).

\clearpage

\section*{}
\label{fig:ext} 

\renewcommand{\figurename}{Extended Data Fig.}
\renewcommand{\tablename}{Extended Data Table}

\begin{table}[ht]
\caption{\bf | Parameters for regions shown in Fig.~\ref{fig:image} obtained with Resolve.}\label{tab:Rsolve}%
\begin{tabular}{c}
~~~~~~~~~~~~~~~~~~~~~~~~~~~~~~~~~~~~~~~~~~~~~~~~~~~~~~~~~~~~~~~~~~~~~~~~~~~~~~~~~~~~~~~~~~~~~~~~~~~~~~~~~~~~~~~~~~~~~~~~~~~~~~~~~~~~~
\end{tabular}
\begin{tabular}{@{}lllll@{}}
\toprule
Region$^*$& Temperature & Fe Abundance & Bulk Velocity$^\dagger$ & Turbulent Velocity \\
 & (keV) & (Solar) & ($\rm km\: s^{-1}$) & ($\rm km\: s^{-1}$) \\
\midrule
Central$^\ddagger$ & $2.81^{+0.18}_{-0.20}$ /$1.41^{+0.20}_{-0.10}$ & $1.65^{+0.07}_{-0.07}$ & $-128^{+6.3}_{-6.3}$ & $117^{+9}_{-9}$ \\
N & $3.10^{+0.15}_{-0.14}$ & $1.31^{+0.12}_{-0.11}$ & $-132^{+14}_{-14}$ & $64^{+27}_{-33}$ \\
NW & $2.91^{+0.15}_{-0.15}$ & $1.34^{+0.15}_{-0.14}$ & $-179^{+21}_{-21}$ & $115^{+29}_{-31}$ \\
W & $2.55^{+0.11}_{-0.12}$ & $1.54^{+0.20}_{-0.17}$ & $-197^{+16}_{-17}$ & $58^{+35}_{-50}$ \\
SW & $2.38^{+0.14}_{-0.09}$ & $2.04^{+0.21}_{-0.20}$ & $-305^{+17}_{-17}$ & $116^{+20}_{-19}$ \\
\botrule
\end{tabular}
Note: For this simultaneous fit, $C$-stat is 23637.16 with 28488 degrees of freedom. As a reference, after grouping the spectra by at least 25 counts in each bin, the chi-square was calculated using the parameters from the spectral fitting results, which was 1868.09 / 1783 in $\chi^2$ / degrees of freedom.\\
*~Errors indicate 1 sigma confidence level and correspond only to statistical errors in the spectral fit.\\
$\dagger$~Heliocentric velocity relative to $3008\rm\: km\: s^{-1}$. The heliocentric velocity of NGC~4696 is $3008\pm 7\rm\: km\: s^{-1}$.\\
$\ddagger$~Two temperature model.
\end{table}


\renewcommand{\arraystretch}{1.3}
\begin{table}[ht]
    \centering
    \caption{{\bf | Properties of individual cavity generations. } }
\begin{tabular}{c}
~~~~~~~~~~~~~~~~~~~~~~~~~~~~~~~~~~~~~~~~~~~~~~~~~~~~~~~~~~~~~~~~~~~~~~~~~~~~~~~~~~~~~~~~~~~~~~~~~~~~~~~~~~~~~~~~~~~~~~~~~~~~~~~~~~~~~
\end{tabular}

    \begin{tabular}{c c c}
    \hline
        Cavities & $E \: (\text{erg})$ & $P_{\text{jet}} \: (\text{erg}\,\text{s}^{-1}$) \\ 
    \hline
        E1--W1 & $1.3_{-0.7}^{+1.4} \times 10^{57}$ & $1.9_{-0.8}^{+1.3} \times 10^{43}$\\
        NE2--SW2 & $2.5_{-1.0}^{+1.5} \times 10^{57}$& $8.5_{-3.7}^{+6.1} \times 10^{42}$\\
        SE3--W3 & $2.0_{-0.8}^{+1.2} \times 10^{57}$ & $4.5_{-1.9}^{+2.7} \times 10^{42}$\\
    \hline
    \end{tabular}
    The cavity names are given in Extended Data Fig.~\ref{fig:cavity}. The total deposited energy $E$, and the corresponding mechanical power of the jet $P_{\text{jet}}$ are shown.
    \label{tab:cavity_powers}
\end{table}

\clearpage

\noindent
\textbf{XRISM Collaboration}
\vspace{3mm}

\noindent
XRISM Collaboration$^{1}$,  
Marc Audard$^{2}$,  
Hisamitsu Awaki$^{3}$,  
Ralf Ballhausen$^{4,5,6}$,  
Aya Bamba$^{7}$,  
Ehud Behar$^{8}$,  
Rozenn Boissay-Malaquin$^{9,5,6}$,  
Laura Brenneman$^{10}$,  
Gregory V.\ Brown$^{11}$,  
Lia Corrales$^{12}$
% ...

\vspace{3mm}
\noindent
$^1$Corresponding Authors: Y. Fujita (y-fujita@tmu.ac.jp) and K. Sato (ksksato@post.kek.jp),  
$^2$Department of Astronomy, University of Geneva, Versoix CH-1290, Switzerland,  
$^3$Department of Physics, Ehime University, Ehime 790-8577, Japan,  
$^4$Department of Astronomy, University of Maryland, College Park, MD 20742, USA,  
$^5$NASA / Goddard Space Flight Center, Greenbelt, MD 20771, USA,  
$^6$Center for Research and Exploration in Space Science and Technology, NASA / GSFC (CRESST II), Greenbelt, MD 20771, USA,  
$^7$Department of Physics, University of Tokyo, Tokyo 113-0033, Japan,  
$^8$Department of Physics, Technion, Technion City, Haifa 3200003, Israel,  
$^9$Center for Space Science and Technology, University of Maryland, Baltimore County (UMBC), Baltimore, MD 21250, USA,  
$^{10}$Center for Astrophysics | Harvard-Smithsonian, MA 02138, USA,  
$^{11}$Lawrence Livermore National Laboratory, CA 94550, USA,  
$^{12}$Department of Astronomy, University of Michigan, MI 48109, USA,  
% ...

\end{document}